# MRI-derived quantification of hepatic vessel-to-volume ratios in chronic liver disease using a deep learning approach

*Original article*

Alexander Herold[1,^], Daniel Sobotka[2,^], Lucian Beer[1], Nina Bastati[1], Sarah Poetter-Lang[1], Michael Weber[1], Thomas Reiberger[3,4,5,6], Mattias Mandorfer[3,4,6], Georg Semmler[3,4,6], Benedikt Simbrunner[3,4,5,6], Barbara D. Wichtmann[7], Sami A. Ba-Ssalamah[1], Michael Trauner[3,6], Ahmed Ba-Ssalamah[1,*], Georg Langs[1,2]

1. Department of Biomedical Imaging and Image-guided Therapy, Medical University of Vienna, Austria.
2. Computational Imaging Research Lab, Department of Biomedical Imaging and Image-guided Therapy, Medical University of Vienna, Austria.
3. Division of Gastroenterology and Hepatology, Department of Medicine III, Medical University of Vienna, Austria.
4. Vienna Hepatic Hemodynamic Lab, Division of Gastroenterology and Hepatology, Department of Medicine III, Medical University of Vienna, Austria
5. Christian Doppler Laboratory for Portal Hypertension and Liver Fibrosis, Medical University of Vienna, Vienna, Austria.
6. Clinical Research Group MOTION, Medical University of Vienna, Vienna, Austria
7. Department of Neuroradiology, University Hospital Bonn, Bonn, Germany.

**Abstract (250 words)**

**Background** We aimed to quantify hepatic vessel volumes across chronic liver disease stages and healthy controls using deep learning-based magnetic resonance imaging (MRI) analysis, and assess correlations with biomarkers for liver (dys)function and fibrosis/portal hypertension.

**Methods** We assessed retrospectively healthy controls, non-advanced and advanced chronic liver disease (ACLD) patients using a 3D U-Net model for hepatic vessel segmentation on portal venous phase gadoxetic acid-enhanced 3-T MRI. Total (TVVR), hepatic (HVVR), and intrahepatic portal vein-to-volume ratios (PVVR) were compared between groups and correlated with: albumin-bilirubin [ALBI] and "model for end-stage liver disease-sodium" [MELD-Na] score) and fibrosis/portal hypertension (Fibrosis-4 [FIB-4] Score, liver stiffness measurement [LSM], hepatic venous pressure gradient [HVPG], platelet count [PLT], and spleen volume.

**Results** We included 197 subjects, aged 54.9 ± 13.8 years (mean ± standard deviation), 111 males (56.3%): 35 healthy controls, 44 non-ACLD, and 118 ACLD patients. TVVR and HVVR were highest in controls (3.9; 2.1), intermediate in non-ACLD (2.8; 1.7), and lowest in ACLD patients (2.3; 1.0) ($p \leq 0.001$). PVVR was reduced in both non-ACLD and ACLD patients (both 1.2) compared to controls (1.7) ($p \leq 0.001$), but showed no difference between CLD groups ($p = 0.999$). HVVR significantly correlated indirectly with FIB-4, ALBI, MELD-Na, LSM, and spleen volume ($\rho$ ranging from -0.27 to -0.40), and directly with PLT ($\rho = 0.36$). TVVR and PVVR showed similar but weaker correlations.

**Conclusions** Deep learning-based hepatic vessel volumetry demonstrated differences between healthy liver and chronic liver disease stages and shows correlations with established markers of disease severity.

**Relevance statement** Hepatic vessel volumetry demonstrates differences between healthy liver and chronic liver disease stages, potentially serving as a non-invasive imaging biomarker.

**Key points**

- Deep learning-based vessel analysis can provide automated quantification of hepatic vascular changes across healthy liver and chronic liver disease stages.
- Automated quantification of hepatic vasculature shows significantly reduced hepatic vascular volume in advanced chronic liver disease compared to non-advanced disease and healthy liver.
- Decreased hepatic vascular volume, particularly in the hepatic venous system, correlates with markers of liver dysfunction, fibrosis, and portal hypertension.

**Abbreviations**

| | |
|---|---|
| ACLD | Advanced chronic liver disease |
| ALBI | Albumin-Bilirubin (Score) |
| CLD | Chronic liver disease |
| FIB-4 | Fibrosis-4 (score) |
| HVPG | Hepatic venous pressure gradient |
| HVVR | Hepatic vein-to-volume ratio |
| IQR | Interquartile range |
| LSM | Liver stiffness measurement |
| MELD-Na | Model for end-stage liver disease-sodium (score) |
| MRI | Magnetic resonance imaging |
| PVVR | Portal vein-to-volume ratio |
| TVVR | Total vessel-to-volume ratio |

# Background

Advanced chronic liver disease (ACLD) is caused by a broad spectrum of underlying entities with viral hepatitis and steatotic liver disease representing the most frequent liver disease etiologies globally. Chronic liver disease (CLD) is characterized by progressive inflammation of the liver parenchyma, resulting in fibrosis and, eventually, cirrhosis [1]. A key pathophysiological consequence is increased intrahepatic vascular resistance due to structural alterations in the hepatic microcirculation and increased vascular tone, ultimately resulting in portal hypertension and related complications [2].

Recent advancements in imaging techniques have contributed to early and noninvasive diagnosis of CLD, providing both morphological and functional information. Techniques such as ultrasound [3] and magnetic resonance elastography are currently utilized for liver fibrosis quantification but require additional equipment and scheduling, thus potentially elevating costs and patient burden [4]. In parallel, gadoxetic acid-enhanced magnetic resonance imaging (MRI) has emerged as a powerful tool in liver imaging, providing both morphological and functional information with prognostic value, including predictions of transplant-free survival [5−7].

While most imaging techniques focus on liver morphology, stiffness, or function, the assessment of hepatic vasculature has received less attention. Recently, contrast-enhanced ultrasound has shown promising results in assessing hepatic microcirculation and noninvasive estimation of hepatic venous pressure gradient [8−11]. Concurrently, advances in artificial intelligence, particularly 3D U-Net architecture, have enabled automated vessel segmentation from various imaging modalities, though primarily focused on preoperative planning [12−17]. However, few studies have applied these techniques to assess hepatic vasculature in CLD [18] representing a significant gap in quantitative vascular analysis for disease assessment. Notably, vascular changes such as reduced hepatic vein diameter may precede other imaging features [19] emphasizing the potential value of quantifying vascular volumetry.

Therefore, this study aimed to quantify liver vessel volumes using a 3D U-Net deep learning model from gadoxetic acid-enhanced MRI. Our main objective was to assess if liver vessel-to-volume ratios – of the total, hepatic venous and portal venous vessels – change with CLD progression, specifically in ACLD compared to non-ACLD patients and a control group. Our secondary goal was to determine correlations between these ratios and established biomarkers for liver (dys)function and fibrosis/portal hypertension.

# Methods

### Patient cohort

This retrospective study was approved by the local institutional review board of the Medical University of Vienna and performed in accordance with the Declaration of Helsinki. The requirement for written informed consent was waived. Patients with histologic or clinical evidence of chronic liver disease who underwent a gadoxetic acid-enhanced liver MRI using a standard examination protocol between 2011 and 2015 were potentially eligible for this study. The patients were selected via a hospital information system database search. Exclusion criteria were: (i) age < 18 years; (ii) known malignancy; (iii) follow-up < 90 days; (iv) cholangiectasis; (v) history of liver transplantation; (vi) Transjugular intrahepatic portosystemic shunt or portal-vein thrombosis; and (vii) poor MRI examination quality. On the basis of the Fibrosis-4 (FIB-4) score (cutoff, 1.75 [20]), which is the diagnostic/prognostic equivalent of 10 kPa for advanced CLD [21, 22], and/or previous or current history of hepatic decompensation, the patients were divided into non-ACLD and ACLD groups. In addition to the two CLD groups, we included a control group consisting of patients who underwent gadoxetic acid-enhanced MRI during the same time interval for evaluation of intraductal papillary mucinous neoplasms. These patients had no history of focal, diffuse, or vascular liver disease based on clinical and radiological assessment, making them suitable as a reference for normal liver vasculature. Fig. 1 shows the patient flowchart.

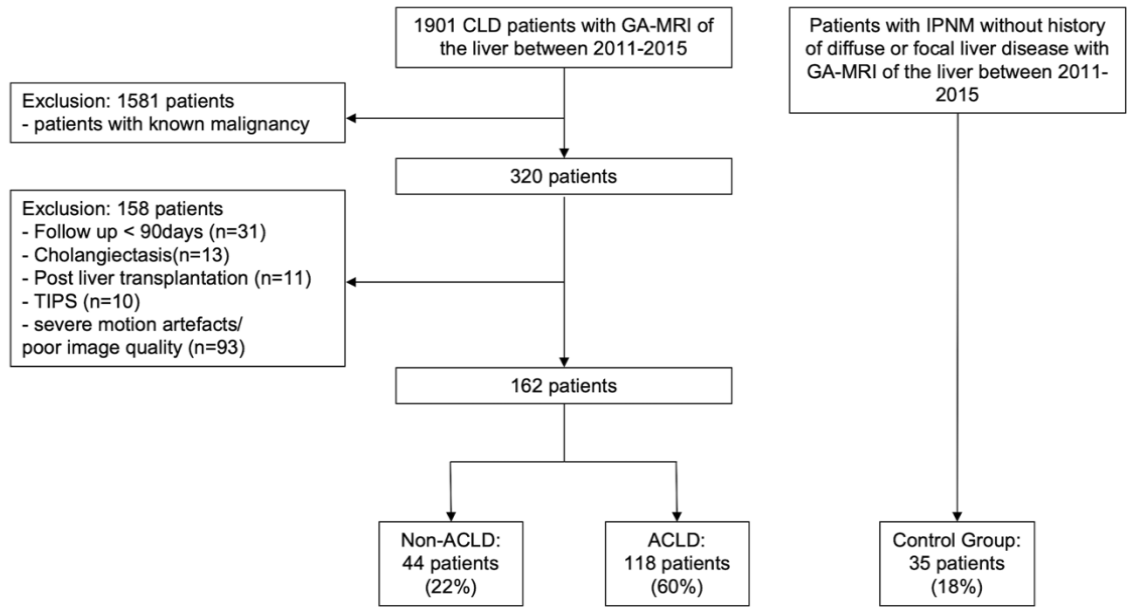

**Fig. 1**. Study flowchart of included patients from the institutional database. *ACLD* Advanced chronic liver disease, *GA* Gadoxetic acid, *IPNM* Intraductal papillary mucinous neoplasms, *MRI* Magnetic resonance imaging, *TIPS* Transhepatic intrajugular portosystemic shunt.

**Clinical Data**

Two authors (G.S. and B.S.) conducted a comprehensive review of patient medical records under the guidance of M.M. and T.R. (board certified hepatologist with extensive experience in the research field of CLD and portal hypertension). To ensure unbiased assessment, the reviewers were blinded to all imaging information during the clinical data collection process. Table 1 presents a summary of the demographic and clinical data.

**MRI protocol**

All MR examinations were performed on a 3-T scanner (Magnetom Trio, A Tim; Siemens Healthcare, Erlangen, Germany). Images were obtained using a combined six-element phased-array coil and a fixed spine coil. MRI examination protocol included unenhanced and dynamic contrast-enhanced, three-dimensional, breath-hold T1-weighted spoiled gradient-echo volumetric sequences, including the hepatobiliary phase, *i.e.*, 20 min after contrast injection, axial diffusion-weighted images, axial in- and opposed-phase T1-weighted images, and conventional T2-weighted images, including coronal cholangiopancreatography and T2-weighted half-Fourier rapid acquisition with relaxation enhancement sequences. A standard dose of gadoxetic acid (0.025 mmol/kg; Primovist/Eovist; Bayer Healthcare, Berlin, Germany) was injected intravenously at a rate of 1.0 mL/s, immediately followed by a 20-mL saline flush. Detailed information on MRI acquisition parameters can be found in the Supplemental Table S1.

**Manual segmentation**

Manual segmentation of hepatic vessels using the ITK-SNAP software (Insight Segmentation and Registration Toolkit - Simpleware Automatic Processor, Version 3.8.0, 2021) was performed by a radiology resident with 3 years of experience on a random subset of 24 patients, including 6 non-ACLD patients, 12 ACLD patients and 6 non-CLD patients. Segmentations were reviewed by a radiologist with over 20 years of experience and revised if necessary. Portal venous phase T1-weighted images were used for annotation, as venous vessel differentiation from other structures, such as bile ducts and arterial vessels, is most practical and more confident in this sequence. Two different labels were used for hepatic veins, including intrahepatic inferior vena cava, as well as the intrahepatic portion of the portal vein and its branches (Fig. 2). Manual segmentation of the liver, spleen, and vessels was performed by the same radiologist on portal venous phase T1-weighted images in all 24 patients. Vessel volumes were normalized to total liver volume to calculate total vessel-to-volume ratio (TVVR), hepatic vein-to-volume ratio (HVVR), and portal vein-to-volume ratio (PVVR).

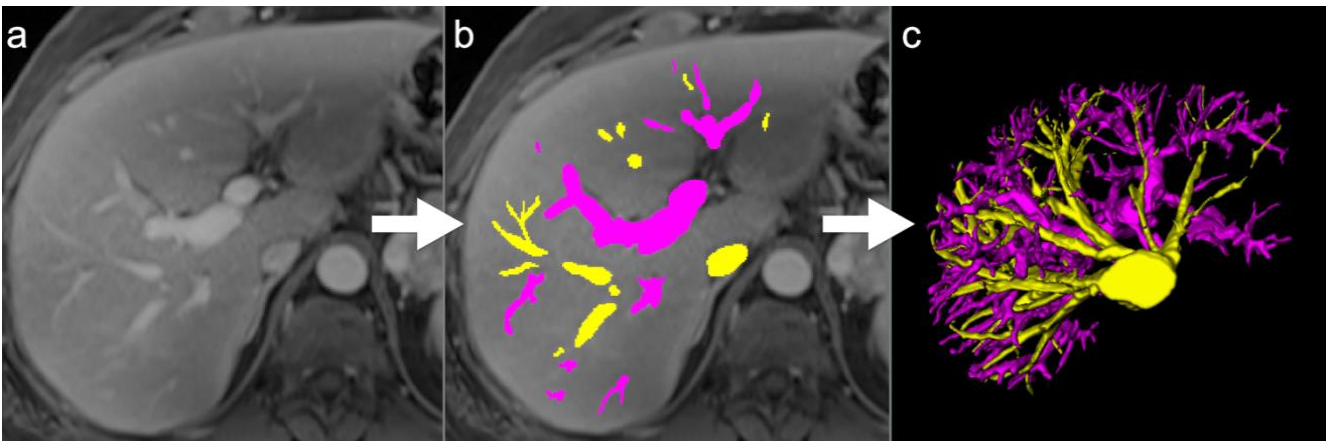

**Fig. 2**. Manual segmentation of liver vessels. (**a**) T1-weighted portal venous images were acquired. (**b**) Manual labeling was performed on these images, with the intrahepatic portions of the portal vein marked in pink and hepatic veins along with the intrahepatic inferior vena cava marked in yellow. (**c**) Three-dimensional reconstruction of the segmented vessel, subsequently utilized for model training.

**3D U-Net architecture and automatic segmentation**

A convolutional neural network with a 3D U-Net architecture [13] was utilized for automated segmentation of liver vessels, hepatic parenchyma and splenic parenchyma (Fig. 3). The model was implemented using PyTorch 1.3.1 with Python 3.7.3, based on the implementation by Wolny et al. [23].

The 3D U-Net comprised of three downsampling blocks and upsampling blocks. Each downsampling block incorporated two $3 \times 3 \times 3$ convolutions, followed by Rectified Linear Units and Group Normalizations, and a 2 × 2 × 2 max pooling operation. The upsampling path, symmetric to the downsampling, utilized nearest neighbor interpolation. Data augmentation included elastic deformation with spline order 3, random flipping, random rotation by 90 degrees, random rotation by ±15 degrees, random contrast, Gaussian noise and Poisson noise. No additional pre- or post-processing was performed.

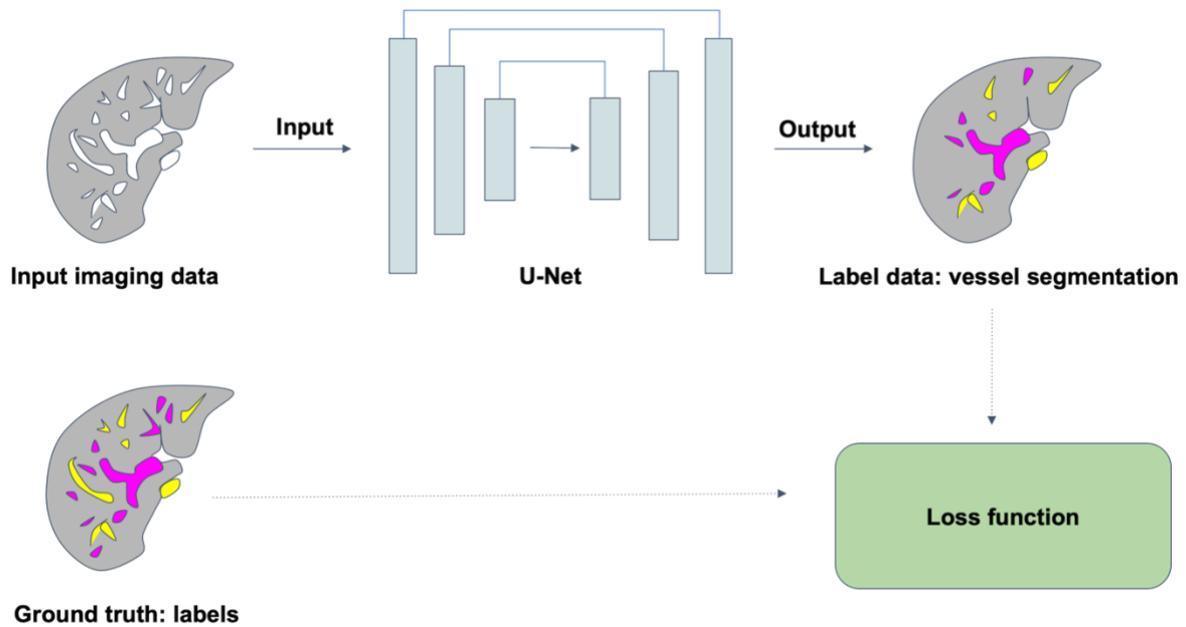

**Fig. 3.** U-Net architecture for liver vessel segmentation. T1-weighted portal venous images and ground truth three-dimensional vessel annotations were used as input, which output three-dimensional vessel segmentations. The model is trained using manually segmented ground truth labels and optimized via a loss function.

The U-Net model was trained on manual segmentations from 24 patients, including 6 non-ACLD, 12 ACLD, and 6 non-CLD patients. Training involved multiple epochs to develop segmentation prediction models for unseen data. 3-fold cross-validation was employed to validate the accuracy of the model. For each cross-validation run, 4 non-ACLD and 8 ACLD patients were excluded from the training set and used for testing. Model optimization was performed using the Adam optimizer [24] with an initial learning rate of 0.001. A batch size of 1 was used with trained patches of size 128 × 96 × 64 and a stride between patches of 32 × 24 × 16. Following testing of the model, liver vessels and parenchyma of the remaining cohort (n=138) were automatically annotated on T1-weighted portal venous MRI sequences.

For each ground truth dataset for each vessel we calculated the minimum distance to the background with Euclidean Distance Transform and use this distance as vessel diameter [25]. Based on these distances, vessel voxels were categorized into four groups: 0−5 mm, 5−10 mm and > 10 mm.

Additionally, for the evaluation of our model on healthy controls, we additionally manually annotated three healthy subjects. These healthy subjects were not used in model training but served as reference standards to calculate Dice scores for the healthy control group. The trained model was then applied to segment vessels in the remaining 32 healthy control patients to calculate vessel-to-volume ratios using identical methodology as for the CLD patients.

**Statistical analysis**

All calculations were performed using SPSS (SPSS Inc., Version 27). Continuous variables were reported as the mean and standard deviation for normally distributed data or, for skewed data, medians with interquartile ranges, respectively. Categorical variables were reported as the number and percentage of patients with specific characteristics. To analyze the differences in vessel-to-volume ratios (TVVR, HVVR, and PVVR) across the control, non-ACLD, and ACLD groups, we used the Kruskal-Wallis test followed by pairwise Mann-Whitney *U* tests with Bonferroni correction. Spearman's rank correlation coefficient ($\rho$) was calculated between vessel-to-volume ratios and various established clinical scores and biochemical markers of liver (dys)function, including albumin, bilirubin, platelet count, albumin-bilirubin (ALBI) score [26], and "model for end-stage liver disease-sodium (MELD-Na) score [27] and direct markers of fibrosis/portal hypertension including FIB-4 score [28], liver stiffness measurement (LSM), hepatic venous pressure gradient (HVPG), platelet count, and spleen volume. The analysis was conducted for the whole CLD cohort ($n$ = 162) and for the two CLD-subgroups, *i.e.*, non-ACLD patients ($n$ = 44 and ACLD patients ($n$ = 118). A *p-value* < 0.05 was considered statistically significant, with Bonferroni correction applied for multiple pairwise comparisons.

Additional subgroup analyses were performed to further characterize differences within the CLD cohort based on clinical disease stage (non-ACLD, compensated ACLD, and decompensated ACLD, following the Baveno VII consensus [21]) and FIB-4 score categories representing low (<1.3), intermediate (1.3-2.67), and high (>2.67) risk for advanced fibrosis [29]. Detailed methodology and results of these analyses are provided in the supplemental material.

For evaluation of segmentation performance, the liver mask segmentation, spleen mask segmentation, and liver vessel segmentation, Dice scores were calculated for the overall segmentation and additionally for each vessel diameter group.

**Declaration of AI and AI-assisted technologies in the writing process**

During the preparation of this work the authors used Claude 3.5 Sonnet AI Assistant (Anthropic, PBC, San Francisco, CA, USA) in order to improve the readability, language and quality of the writing. After using this service, the authors reviewed and edited the content as needed and take full responsibility for the content of the publication.

## Results

A total of 197 patients were included:
- 162 patients with CLD, aged 53.3 ± 13.9 years (mean ± standard deviation), 99 males (61.1%), specifically 44 non-ACLD and 118 ACLD patients;
- 35 patients without liver disease as controls, aged 62.5 ± 11.7 years, 12 males (34.3%).

**Table 1.** Patient characteristics

| Characteristic | Control group (n = 35) | Non-ACLD (n = 44) | ACLD (n = 118) | p-value |
|---|---|---|---|---|
| **Age, mean ± standard deviation** | 62.5 ± 11.7 | 41.4 ± 12.6 | 57.7 ± 12.3 | < 0.001 |
| **Sex (male), n (%)** | 12 (34.3%) | 22 (50.0%) | 77 (65.3%) | 0.003 |
| **Body mass index** | 24.0 (21.5−27.6) | 23.8 (21.0−27.4) | 24.8 (22.5−28.2) | 0.523 |
| **Etiology, n (%)** | | | | |
|   MetALD/ALD | − | 4 (9.1%) | 26 (22.0%) | − |
|   Hepatitis C virus | − | 5 (11.4%) | 35 (29.7%) | − |
|   Hepatitis B virus | − | 6 (13.6%) | 9 (7.6%) | − |
|   MASLD/MASH | − | 4 (9.1%) | 8 (6.8%) | − |
|   PSC | − | 10 (22.7%) | 6 (5.1%) | − |
|   AIH | − | 2 (4.5%) | 5 (4.2%) | − |
|   PBC | − | 1 (2.3%) | 7 (5.9%) | − |
|   SCC | − | 1 (2.3) | 0 (0.0) | − |
|   Metabolic | − | 0 (0.0%) | 2 (1.7%) | − |
|   Cryptogenic | − | 8 (18.2%) | 20 (16.9) | − |
| **Albumin (g/dL)** | − | 42.2 (37.2−45.0) | 39.0 (33.0−42.3) | = 0.007 |
| **Bilirubin (mg/dL)** | − | 0.5 (0.4−1.0) | 1.0 (0.6−2.1) | < 0.001 |
| **PTL (Giga/L)** | − | 295.5 (222.5−355.0) | 130.5 (81.2−174.8) | < 0.001 |
| **Albumin-bilirubin score** | − | -2.9 (-3.2−-2.5) | -2.4 (-2.9−-1.9) | < 0.001 |
| **Fibrosis-4 score** | − | 0.8 (0.6−1.0) | 3.5 (2.2−6.1) | < 0.001 |
| **MELD-Na Score** | − | 6.6 (6.0−10.9) | 8.3 (6.0−14.9) | = 0.043 |
| **Liver stiffness measurement (kPa)** | − | 7.9 (6.1−11.8) (n=14) | 25.7 (13.1−37.4) (n=51) | < 0.001 |
| **HVPG (mmHg)** | − | − | 16.0 (13.0−21.8) (n=50) | − |
| **Spleen Volume (cc)** | 130.2 (69.9−218.9) | 232.1 (132.3−330.7) | 405.1 (210.6−695.5) | < 0.001 |
| **Liver Volume (cc)** | 1289.5 (1113.9−1663.0 | 1712.2 (1407.0−2008.3) | 1758.2 (1346.4−2219.7) | < 0.001 |

**Table 1:** Patient Characteristics. All values are presented as median (interquartile range) unless indicated otherwise. *MASLD* Metabolic dysfunction-associated steatotic liver disease, *MASH* Metabolic Dysfunction-Associated Steatohepatitis, *MetALD/ALD* Patients with MASLD who consume greater amounts of alcohol per week/Alcoholic Liver Disease, *FIB-4* Fibrosis-4, *MELD-Na* Model for End-Stage Liver Disease – Sodium, *HVPG* hepatic venous pressure gradient.

In the overall CLD cohort, viral hepatitis (34.0%, *n* = 55) and alcoholic liver disease (18.5%, *n* = 30) were the most common etiologies. However, the distribution differed between groups: ACLD patients showed predominantly viral hepatitis (37.3%, n = 44) and alcoholic liver disease (22.0%, *n* = 26), while non-ACLD patients had a higher prevalence of primary sclerosing cholangitis (22.7%, *n* = 10) and cryptogenic liver disease (18.2%, *n* = 8). Detailed demographic information and etiology distribution for each group can be found in Table 1.

**Comparison of vessel-to-volume ratios between subgroups**

Significant differences in liver vessel volume metrics were observed across healthy controls, non-ACLD, and ACLD patients (Figs. 4 and 5, Table 2a,b). Kruskal-Wallis testing demonstrated significant differences across all three groups for TVVR, HVVR, and PVVR (all *p* < 0.001). The TVVR was highest in healthy controls (median: 3.9, interquartile range [IQR]: 3.3−4.5), lower in non-ACLD patients (median: 2.8, IQR: 2.3-3.8), and lowest in ACLD patients (median: 2.3, IQR: 1.6-3.0).

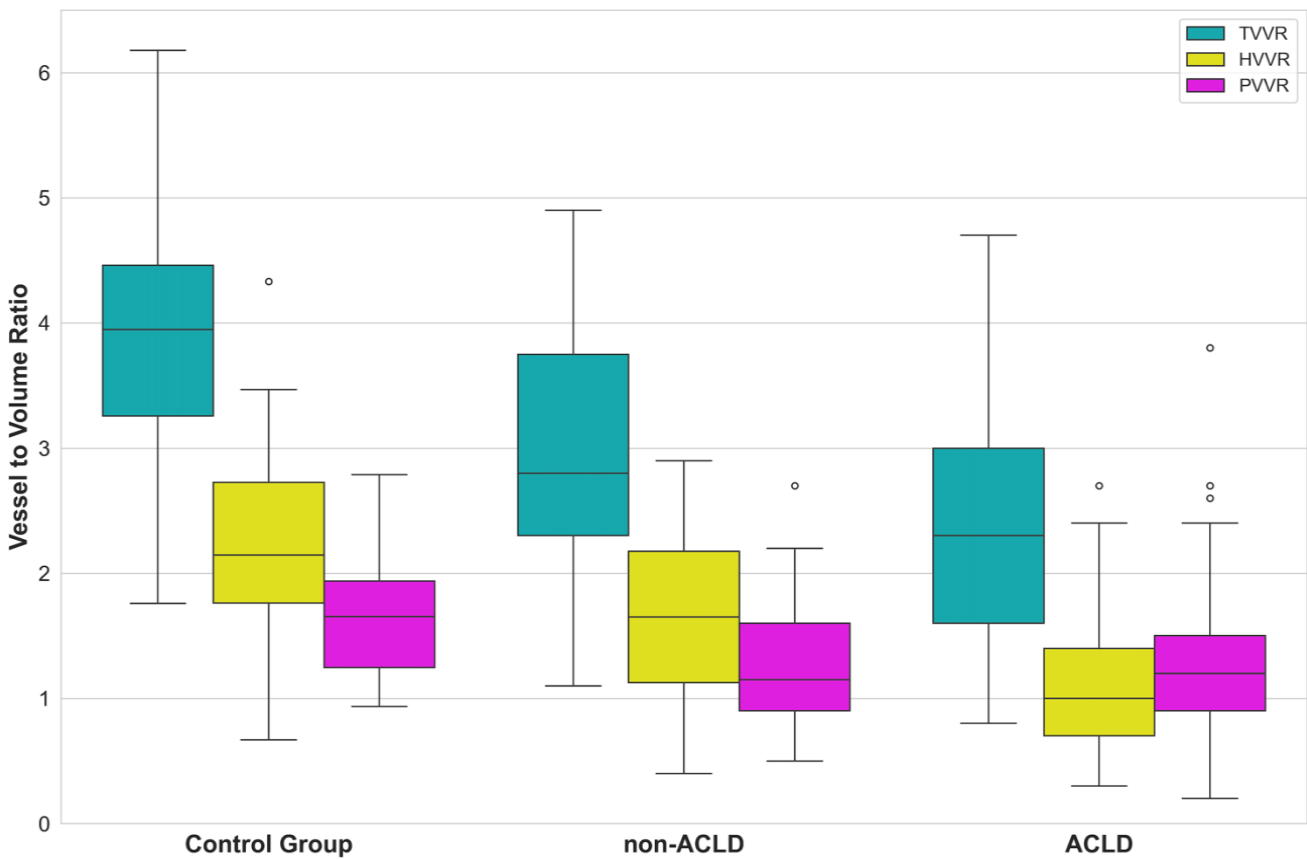

**Fig. 4.** Comparison of vessel-to-volume ratios across the control group, non-advanced chronic liver disease (non-ACLD), and advanced chronic liver disease (ACLD) patients. Boxplots show the distribution of total vessel-to-volume ratio (TVVR), hepatic vein-to-volume ratio (HVVR), and portal vein-to-volume ratio (PVVR) across all three groups. Progressive decreases in TVVR and HVVR are observed from controls to non-ACLD to ACLD, while PVVR shows significant reduction only between healthy controls and disease groups.

Pairwise comparisons with Bonferroni correction revealed significant differences of 28.2% between controls and non-ACLD (*p* < 0.001), 17.9% between non-ACLD and ACLD (*p* = 0.006), and a total difference of 41.0% between controls and ACLD ($p < 0.001$). These differences were primarily driven by HVVR, which showed distinct differences across groups, with the highest values in healthy controls (median: 2.1, IQR: 1.8-2.7), intermediate values in non-ACLD patients (median: 1.7, IQR: 1.1-2.2), and lowest values in ACLD patients (median: 1.0, IQR: 0.7-1.4). These differences represented reductions of 19.0% between controls and non-ACLD ($p = 0.005$), 41.2% between non-ACLD and ACLD ($p < 0.001$), and 52.4% between controls and ACLD ($p < 0.001$). In contrast, PVVR showed a significant 29.4% difference between healthy controls (median: 1.7, IQR: 1.2-1.9) and non-ACLD patients (median: 1.2, IQR: 0.9-1.6) ($p = 0.010$) and a similar 29.4% difference between controls and ACLD patients (median: 1.2, IQR: 0.9-1.5) ($p < 0.001$), but no significant difference between non-ACLD and ACLD groups ($p = 0.999$).

Additional subgroup analyses of the CLD group revealed no significant differences in vessel-to-volume ratios between ACLD subgroups, nor between intermediate and high risk groups for advanced liver fibrosis based on FIB-4 score (Tables S2-S3 and Figures S1-S2 of the supplemental material).

**Table 2a.** Vessel-to-volume ratios by group

| Parameter | Control group ($n = 35$) | Non-ACLD ($n = 44$) | ACLD ($n = 118$) | $p$-value* |
|---|---|---|---|---|
| **TVVR** | 3.9 (3.3−4.5) | 2.8 (2.3−3.8) | 2.3 (1.6−3.0) | < 0.001 |
| **HVVR** | 2.1 (1.8−2.7) | 1.7 (1.1−2.2) | 1.0 (0.7−1.4) | < 0.001 |
| **PVVR** | 1.7 (1.2−1.9) | 1.2 (0.9−1.6) | 1.2 (0.9−1.5) | < 0.001 |

Data are presented as median (interquartile range). *Kruskal-Wallis test. *ACLD* Advanced chronic liver disease, *HVVR* Hepatic vein-to-volume ratio, *PVVR* Portal vein-to-volume ratio, *TVVR* Total vessel-to-volume ratio.

**Table 2b.** Pairwise comparisons

| Parameter | Percentage decreases | | | $p$-values* | | |
|---|---|---|---|---|---|---|
| | Control → Non-ACLD | Control → ACLD | Non-ACLD → ACLD | Control versus Non-ACLD | Control versus ACLD | Non-ACLD versus ACLD |
| **TVVR** | -28.2% | -41.0% | -17.9% | < 0.001 | < 0.001 | 0.006 |
| **HVVR** | -19.0% | -52.4% | -41.2% | =0.005 | < 0.001 | < 0.001 |
| **PVVR** | -29.4% | -29.4% | 0.0% | =0.010 | < 0.001 | =0.999 |

Percent decreases were calculated from median values. * Pairwise Mann-Whitney *U* tests with Bonferroni correction for multiple comparisons. *ACLD* Advanced chronic liver disease, *HVVR* Hepatic vein-to-volume ratio, *PVVR* Portal vein-to-volume ratio, *TVVR* Total vessel-to-volume ratio.

**Spearman correlation of vessel-to-volume ratios and clinical parameters**

In the overall CLD cohort, TVVR showed negative correlations with spleen volume (ρ = -0.24, *p* = 0.003), HVPG (ρ = -0.33, *p* = 0.020), LSM (ρ = -0.28, *p* = 0.030), FIB-4 (ρ = -0.27, *p* = 0.004), ALBI (ρ = -0.36, *p* < 0.001), MELD-Na (ρ = -0.27, *p* < 0.001) and bilirubin (ρ = -0.31, *p* < 0.001), and positive correlations with albumin (ρ = 0.31, *p* < 0.001) and platelet count (ρ = 0.24, *p* = 0.005) (Fig. 6). HVVR correlated negatively with spleen volume (ρ = -0.30, *p* < 0.001), LSM (ρ = -0.40, *p* < 0.001), FIB-4 (ρ = -0.39, *p* < 0.001), ALBI (ρ = -0.33, *p* < 0.001), MELD-Na (ρ = -0.27, *p* < 0.001), bilirubin (ρ = -0.33, *p* < 0.001), and positively with albumin (ρ = 0.27, *p* < 0.001) and platelet count (ρ = 0.36, *p* < 0.001). PVVR correlated negatively with HVPG (ρ = -0.28, *p* = 0.040), ALBI (ρ = -0.28, *p* < 0.001), MELD-Na (ρ = -0.18, *p* = 0.030) and bilirubin (ρ = -0.23, *p* < 0.01), and positively with albumin (ρ = 0.25, *p* = 0.003).

In ACLD patients, TVVR correlated negatively with HVPG (ρ = -0.33, *p* = 0.020), ALBI (ρ = -0.31, *p* = 0.005), MELD-Na (ρ = -0.3, *p* = 0.009) and bilirubin (ρ = -0.29, *p* = 0.07), and positively with albumin (ρ = 0.26, *p* = 0.005). HVVR correlated negatively with LSM (ρ = -0.29, *p* = 0.040), ALBI (ρ = -0.26, *p* = 0.008), MELD-Na (ρ = -0.34, *p* < 0.001), bilirubin (ρ = -0.33, *p* < 0.001), and positively with albumin (ρ = 0.18, *p* = 0.040). PVVR correlated negatively with HVPG (ρ = -0.29, *p* = 0.040), ALBI (ρ = -0.3, *p* < 0.003) and bilirubin (ρ = -0.19, *p* = 0.040), and positively with albumin (ρ = 0.28, *p* = 0.006).

In non-ACLD patients, only PVVR showed a significant correlation with bilirubin (ρ = -0.34, *p* = 0.022).

**Liver and vessel segmentation cross-validation**

The liver mask segmentation achieved a Dice score of 0.97 for both non-ACLD and ACLD groups. The spleen mask segmentation achieved a Dice score of 0.93 for non-ACLD and 0.97 for ACLD, respectively. For liver vessel segmentation, Dice scores of 0.63 and 0.67 were obtained for the non-ACLD and ACLD groups, respectively. The model's generalizability was assessed by evaluating the performance of liver vessel segmentation in the healthy control group. The model achieved a Dice score of 0.71 for the control group. Further analysis of segmentation accuracy based on vessel size was then performed. Large vessels (> 10 mm in diameter) were segmented with high to moderate accuracy, yielding Dice scores of 0.65 for the non-ACLD group, and 0.76 for the ACLD group. Medium-sized vessels (5−10 mm) showed similar accuracy, with Dice scores of 0.70 for non-ACLD, and 0.74 for ACLD groups, respectively. The model's performance was lower for small vessels (< 5 mm), resulting in Dice scores of 0.47 for the non-ACLD group, and 0.50 for the ACLD group.

For the non-ACLD datasets, 51.9% of vessels fell within the 0−5-mm group, 31.3% within the 5−10-mm group, and 16.74% within the > 10-mm group. For the ACLD datasets, 42.63% of vessels were in the 0−5-mm group, 32.7% in the 5−10-mm group, 19.66% in the >15-mm group.

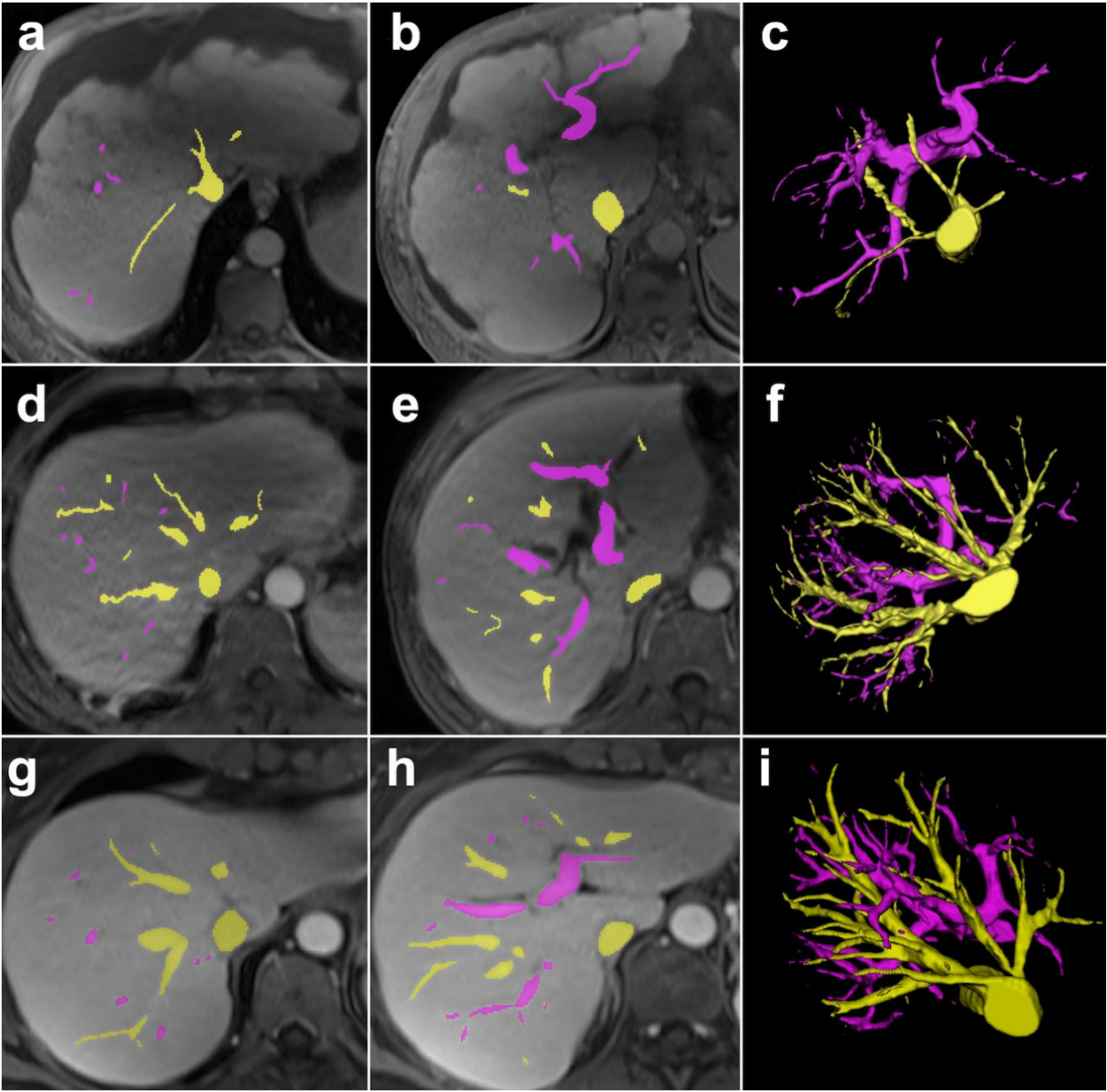

**Fig. 5.** Liver vessel volume in an advanced chronic liver disease (ACLD) patient (top row, **a-c**) *versus* a non-ACLD patient (mid row, **d-f**) *versus* a control group patient (bottom row, **g-i**) with median total vessel-to-volume-ratio of 1.8 *versus* 2.8 *versus* 4.5, hepatic vein-to-volume ratio of 0.8 *versus* 1.7 *versus* 2.7 and portal vein to volume ratio of 1.0 *versus* 1.1 *versus* 1.8, respectively. Hepatic veins (yellow) demonstrate substantially lower volume in the ACLD patient (**a**) compared to the non-ACLD patient (**d**) and control group patient (**g**). Portal veins (pink) demonstrate less pronounced differences (**b**, **e**, **h**). Three-dimensional reconstructions highlight overall vessel volume disparity (**c**, **f**, **i**).

## Discussion

In this study, we demonstrated the application of 3D U-Net deep learning-based segmentation to quantify liver vasculature and examine liver vessel-to-volume ratios across different groups, from healthy liver to advanced chronic liver disease. We observed marked differences in vessel volumes between groups, with total vessel-to-volume ratio (TVVR) showing significant reductions of 28.2% in controls and non-ACLD patients and a further 17.9% in ACLD patients. These differences were primarily driven by reductions in hepatic vein-to-volume ratio (HVVR), which was 19.0% lower in non-ACLD than controls and 41.2% lower in ACLD than non-ACLD patients. In contrast, portal vein-to-volume ratio (PVVR) showed a significant 29.4% reduction from controls to non-ACLD but remained stable between non-ACLD and ACLD groups. Moreover, we observed significant correlations between these vessel-to-volume ratios and established markers of liver dysfunction and fibrosis/portal hypertension.

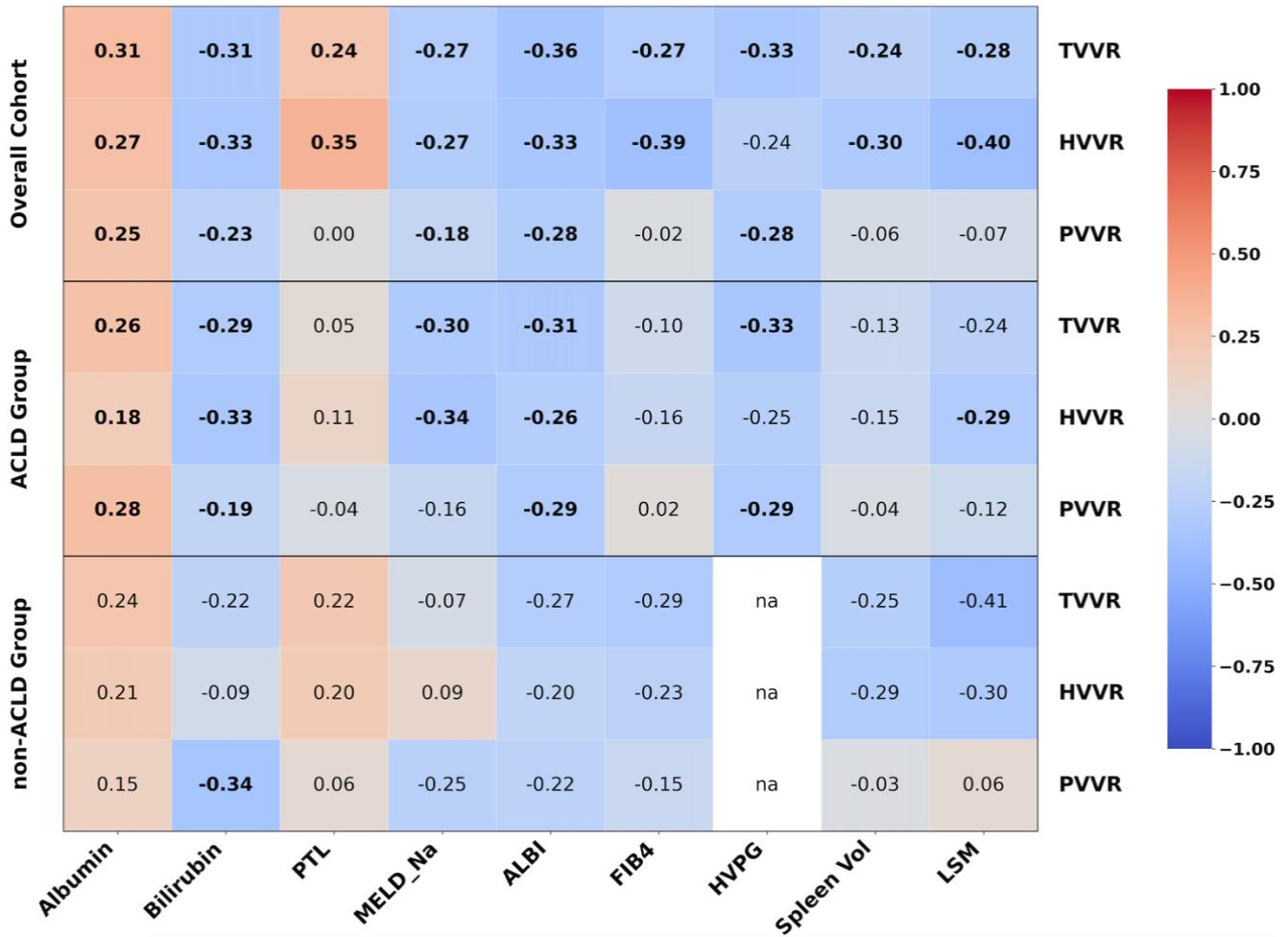

**Fig. 6**. Correlation map of liver vessel-to-volume ratios and biomarkers for liver (dys)function and fibrosis/portal hypertension. Values represent Spearman's correlation coefficient (ρ). Statistically significant correlations ($p < 0.05$) are highlighted in bold. *ACLD* Advanced chronic liver disease, *HVVR* Hepatic vein-to-volume ratio, *PVVR* Portal vein-to-volume ratio, *TVVR* Total vessel-to-volume ratio.

The observed decrease in HVVR in ACLD patients aligns with a study by Zhang et al., who analyzed changes in vascular morphology on MRI in patients with cirrhosis compared to healthy liver donors. The group demonstrated that the main hepatic vein diameter in the cirrhotic liver was significantly smaller than in non-cirrhotic livers, with a sensitivity and specificity of 88% and 85% using a cutoff of 7 mm for the right hepatic vein [19]. The relationship between cirrhosis and alterations in hepatic vessel size is complex and likely involves a combination of both parenchymal and vascular/hemodynamic changes. Primarily, the reduced vessel volume may be attributed to vascular compression related to increased liver stiffness due to fibrosis and nodule formation, which may exert external pressure on the hepatic vessels. Concurrently, vascular adaptations occur in response to the altered hemodynamics in cirrhosis. While portal hypertension generally leads to increased portal vein flow and the development of collaterals, the hepatic veins may experience reduced blood flow due to portosystemic shunting. This shunting diverts blood away from the liver, potentially contributing to a decrease in hepatic vein volume. However, we should note that the pressure in the hepatic veins, which corresponds to that in the inferior vena cava, does not change significantly in cirrhosis.

In case of concurrent congestive heart failure, the presence of an enlarged vena cava and hepatic veins can lead to misleading results in studies trying to associate the size of hepatic veins with cirrhosis, potentially producing false negatives. Therefore, patients with concurrent heart failure may show a normal hepatic vein diameter even when cirrhosis is present [29, 30].

While changes in vascular size might often be difficult to appreciate on conventional computed tomography and MRI, especially in non-advanced CLD patients with a lack of other cirrhotic imaging features, volumetric deep learning analysis, as performed in our study, might enable the detection of cumulative vessel volume loss. Notably, our findings demonstrate different patterns of vascular changes between hepatic and portal venous systems. HVVR showed progressive differences across all groups, with additional subgroup analyses (Supplemental Material) suggesting that these changes may occur early in disease progression and potentially stabilize thereafter, as we observed significant differences as we observed significant differences between non-ACLD and compensated ACLD but not between compensated and decompensated stages, and between low and intermediate risk but not between intermediate and high risk for advanced liver fibrosis based on FIB-4 score. In contrast, PVVR significantly decreased from healthy controls to non-ACLD patients but showed no further difference between non-ACLD and ACLD groups, nor within ACLD subgroups or FIB-4 category. This pattern likely reflects the complex hemodynamic adaptations in portal hypertension. We recognize that a substantial portion of portal venous volume is comprised not only of the main portal vein and its right and left branches, but also numerous smaller intrahepatic branches. The initial reduction in PVVR from healthy to non-ACLD may represent early compression of these smaller portal branches due to developing fibrosis, while the subsequent stabilization between non-ACLD and ACLD could reflect competing forces: continued compression of smaller intrahepatic branches by advancing fibrosis on one hand, and compensatory dilation of the main portal vein and larger branches due to increasing portal pressure on the other [31]. This interpretation aligns with previous studies showing that extrahepatic portal venous diameter is significantly larger in patients with CLD and cirrhosis compared to healthy controls [32-34], while our measurements focused on the entire intrahepatic portal venous system including smaller branches, which might respond differently to disease progression than the main portal vein.

Our study demonstrated weak to moderate correlations between vessel-to-volume ratios and established clinical scores and biochemical markers of liver dysfunction. Specifically, the consistent trends of correlation observed between vessel-to-volume ratios and biomarkers indicating worsening synthetic dysfunction (lower albumin) [35], excretory dysfunction (higher bilirubin) [36], clinical scores of liver impairment and independent predictors of survival (ALBI- [37] and MELD-Na scores [27]) suggest reduced intrahepatic vascular volume relative to parenchymal volume with advancing disease. Negative correlations were observed between vessel-to-volume ratios and the FIB-4 Score [38], LSM, HVPG and spleen volume, further extending our findings to markers of liver fibrosis and portal hypertension. The negative correlation between PVVR and HVPG may seem counterintuitive, as higher portal pressure typically correlates with increased portal vein diameter. This finding might be explained by our segmentation method, which only included the intrahepatic portion of portal vein, not capturing changes in the extrahepatic portal vein.

In addition, the positive correlations seen with platelet count also align with the multifactorial causes of thrombocytopenia in chronic liver disease, including decreased platelet production [39], hypersplenism related to portal hypertension [40] and increased platelet destruction due to decreased levels of A disintegrin-like and "metalloprotease with thrombospondin type 1 motif 13"−ADAMTS13 [41] and immunologically mediated destruction [42]. Of note, the HVVR and, therefore, TVVR correlated more strongly and consistently with clinical biomarkers than the PVVR. As the vessel-to-volume ratios capture vascular changes that accompany the spectrum of ACLD and resulting portal hypertension, these findings suggest their ability to non-invasively support the diagnosis of ACLD.

Several limitations of this study need to be acknowledged. First, the retrospective nature and referral patterns to our tertiary center may introduce selection bias. However, our cohort represents a broad spectrum of liver diseases, including viral hepatitis, metabolic, cholestatic, autoimmune and cryptogenic liver disease, reflecting real-world patient distribution at a tertiary referral center. In addition, gadoxetic acid-enhanced MRI liver examinations are routinely performed at our institution for assessment of CLD and focal liver lesions. Therefore, selection bias was less likely to impact our cohort. Second, histopathologic confirmation of etiology and fibrosis stage was not available in most cases, as biopsy is often not needed for establishing the etiology of liver disease and also is no longer standard clinical practice for staging fibrosis. With inclusion of more sophisticated non-invasive tests such as vibration-controlled transient elastography (VCTE), we provided a more accurate estimate of liver fibrosis and portal hypertension [43], however, considering the inclusion period, they were less broadly available/used, as compared to today's clinical practice. While histology would further validate imaging findings, our patient distribution reflects the reality of the clinical routine. Third, our deep learning segmentation model was trained on a relatively small dataset of 24 patients. Additional manual segmentation data could help refine model performance and enhance the reliability of vessel volume quantification. Furthermore, our model was validated only on gadoxetic acid-enhanced 3-T MRI data. Lower field strengths like 1.5 T typically provide reduced signal-to-noise ratio and spatial resolution, which could affect vessel delineation, particularly of smaller branches. Different contrast agents might alter vessel-to-liver contrast and enhancement patterns, potentially impacting segmentation performance. Application at different field strengths or with different contrast agents would require additional validation to ensure robust vessel segmentation performance.

Fourth, our deep learning segmentation model was trained on a relatively small dataset of 24 patients, and showed lower performance for small vessels (< 5 mm). While additional manual segmentation data could help refine model performance and enhance the reliability of vessel volume quantification, the consistent performance across patient groups is reassuring. To validate the model, we manually annotated three healthy subjects (not used for training) to serve as a reference standard, then applied our model to the entire healthy control cohort. The resulting Dice score for the healthy group were comparable to for non-ACLD and ACLD groups, demonstrating consistent performance across different patient populations.

Fifth, we did not systematically assess cardiac function in our cohort. While we did not include patients with congestive hepatopathy/cardiac cirrhosis as their primary hepatic disease, we cannot rule out the presence of concurrent cardiac dysfunction in some patients, especially those secondary to hyperdynamic circulation and cirrhotic cardiomyopathy. This could have led to increased hepatic vein volumes, which would have worked against our observation of reduced hepatic vein volumes in advanced liver disease. Notably, since right heart failure due to hyperdynamic circulation would predominantly affect patients with ACLD, this would make our finding of lower hepatic vein volumes in this group even more convincing. Future studies should include systematic cardiac evaluation to better understand these relationships.

Lastly, as a single-center retrospective analysis, our findings require external validation in larger, multicenter contemporary CLD (*i.e.*, higher proportion of steatotic liver disease) cohorts to confirm generalizability across populations and clinical settings.

In conclusion, our study demonstrates that hepatic vessel volumes exhibit distinct differences between healthy liver and different stages of chronic liver disease that can be quantitatively measured using deep learning-based gadoxetic acid-enhanced 3-T MRI analysis. Computational analysis of vessel-to-liver volume ratios allows non-invasive detection of vascular changes that could serve as an additional imaging biomarker for the staging of CLD.

# Supplemental Material

**Subgroup Analysis of Hepatic Vessel-to-Volume Ratios:**

To further analyze the chronic liver disease group and explore the relationship between progressive liver disease severity and vessel-to-volume ratios, we performed two additional subgroup analyses beyond the main three-group comparison (healthy controls, non-ACLD, and ACLD) presented in the manuscript.

First, we stratified the chronic liver disease patients into three categories: non-ACLD (n=44), compensated ACLD (cACLD, n=74), and decompensated ACLD (dACLD, n=44). This classification aligns with the Baveno VII consensus [21] that defines cACLD as patients with untreated/active chronic liver disease at risk of having clinically significant portal hypertension, and consequently, at risk of decompensation and liver-related death.

As shown in Table S2, when dividing patients into non-ACLD, cACLD, and dACLD groups, we observed significant differences in TVVR (p=0.0017) and HVVR (p<0.0001) across groups, while PVVR remained stable (p=0.4021). Post-hoc pairwise comparisons revealed that TVVR significantly differed between non-ACLD and cACLD (p=0.033) as well as between non-ACLD and dACLD (p=0.001), but not between cACLD and dACLD (p=0.443). Similarly, HVVR showed significant differences between non-ACLD and cACLD (p<0.001) and between non-ACLD and dACLD (p<0.001), but not between cACLD and dACLD (p=1.000). These findings are visually represented in Figure S1.

Second, we stratified patients based on FIB-4 score cutoffs (<1.3, 1.3-2.67, >2.67) representing low, intermediate, and high risk for advanced fibrosis, respectively following the established thresholds as mentioned by Sterling et al. [29].

Similarly, our FIB-4 score-based stratification (Table S3) demonstrated significant differences in TVVR (p=0.0072) and HVVR (p<0.0001) across the three groups, with no significant differences in PVVR (p=0.4730). Pairwise comparisons showed significant differences in TVVR between FIB-4 <1.3 and 1.3-2.67 (p=0.026) and between FIB-4 <1.3 and >2.67 (p=0.010). For HVVR, significant differences were observed between FIB-4 <1.3 and 1.3-2.67 (p=0.001) and between FIB-4 <1.3 and >2.67 (p<0.001). These findings are illustrated in Figure S2.

Table-S1. MRI protocol

| Sequence | Section Thickness (mm) | TR (msec) | TE (msec) | FOV (mm) | Phase Direction | Flip Angle |
|---|---|---|---|---|---|---|

| Sequence | | | | | | |
|---|---|---|---|---|---|---|
| GRE-T1 (flash 2D) in-phase | 5 | 130 | 2.5 | 350 | AP | 70 |
| GRE-T1 (flash 2D) opp-phase | 5 | 131 | 3.7 | 350 | AP | 70 |
| T1 VIBE SPAIR axial | 1.7 | 2.7 | 1.0 | 430 | AP | 13 |
| T1 VIBE SPAIR coronal | 2 | 2.6 | 0.9 | 500 | RL | 13 |
| T2 Haste coronal | 4.5 | 805 | 76 | 450 | RL | 141 |
| DWI axial TSE-EP | 6 | 1700 | 73 | 380 | AP | — |
| T2 MRCP 2D-HASTE | 45 | 5500 | 454 | 380 | RL | 180 |
| T2 MRCP 3D-SPACE | 0.9 | 2400 | 707 | 350 | RL | variable |
| T2 Haste axial fs. | 5 | 1800 | 150 | 400 | AP | 150 |

MRI = magnet resonance imaging; 2D = Two-dimensional; FOV = field of view; fs = fat saturation; GRE = gradient echo; SPAIR = spectral attenuated inversion recovery; TE = echo time; TR = repetition time; TSE = turbo spin echo; VIBE = volumetric interpolated breath-hold examination; MRCP = magnetic resonance cholangiopancreatography; HASTE = Half-Fourier-Acquired Single-shot Turbo spin Echo.

Table-S2. Vessel-to-volume ratios by CLD Group

| Parameter | non-ACLD (n=44) | cACLD (n=74) | dACLD (n=44) | p-value* | Post-hoc comparisons† |
|---|---|---|---|---|---|
| TVVR | 2.8 (2.3-3.8) | 2.3 (1.8-3.3) | 2.3 (1.5-3.0) | 0.0017 | a: p=0.033, b: p=0.001, c: p=0.443 |
| HVVR | 1.7 (1.1-2.2) | 1.0 (0.7-1.5) | 0.9 (0.7-1.4) | <0.0001 | a: p<0.001, b: p<0.001, c: p=1.000 |
| PVVR | 1.2 (0.9-1.6) | 1.3 (0.9-1.6) | 1.20 (0.8-1.6) | 0.4021 | a: p=1.000, b: p=1.000, c: p=0.584 |

Comparison of vessel-to-volume ratios across clinical disease stages. The Kruskal-Wallis test was used to assess overall differences between groups, followed by pairwise Mann-Whitney U tests with Bonferroni correction for multiple comparisons. *Kruskal-Wallis test; †Post-hoc pairwise comparisons (Dunn's test with Bonferroni correction): a: non-ACLD vs cACLD, b: non-ACLD vs dACLD, c: cACLD vs dACLD. TVVR = total vessel-to-volume ratio; HVVR = hepatic vein-to-volume ratio; PVVR = portal vein-to-volume ratio; non-ACLD = non-advanced chronic liver disease; cACLD = compensated advanced chronic liver disease; dACLD = decompensated advanced chronic liver disease.

Table-S3. Vessel-to-volume ratios by FIB-4-Score

| Parameter | FIB4 <1.3 (n=46) | FIB4 1.3-2.67 (n=47) | FIB4 >2.67 (n=69) | p-value* | Post-hoc comparisons† |
|---|---|---|---|---|---|
| TVVR | 2.8 (2.4-3.8) | 2.3 (1.7-3.2) | 2.3 (1.6-3.0) | 0.0072 | a: p=0.026, b: p=0.010, c: p=1.000 |
| HVVR | 1.7 (1.3-2.2) | 0.9 (0.7-1.6) | 1.0 (0.7-1.4) | <0.0001 | a: p=0.001, b: p<0.001, c: p=1.000 |
| PVVR | 1.2 (0.9-1.7) | 1.1 (0.9-1.4) | 1.4 (0.9-1.8) | 0.4730 | a: p=1.000, b: p=1.000, c: p=0.665 |

Comparison of vessel-to-volume ratios stratified by FIB-4 score categories. The Kruskal-Wallis test was used to assess overall differences between groups, followed by pairwise Mann-Whitney U tests with Bonferroni correction for multiple comparisons. Significant p-values are highlighted in bold. *Kruskal-Wallis test; †Post-hoc pairwise comparisons (Dunn's test with Bonferroni correction): a: <1.3 vs 1.3-2.67, b: <1.3 vs >2.67, c: 1.3-2.67 vs >2.67; TVVR = total vessel-to-volume ratio; HVVR = hepatic vein-to-volume ratio; PVVR = portal vein-to-volume ratio.

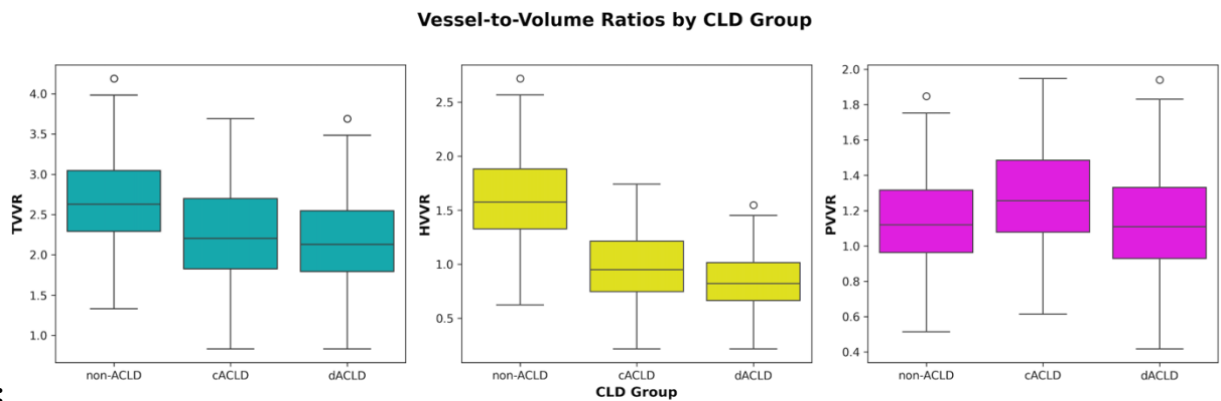

**Fig. S1:**

Fig. S1: Distribution of vessel-to-volume ratios across clinical disease stages. Box plots show total vessel-to-volume ratio (TVVR), hepatic vein-to-volume ratio (HVVR), and portal vein-to-volume ratio (PVVR) across non-advanced chronic liver disease (non-ACLD), compensated advanced chronic liver disease (cACLD), and decompensated advanced chronic liver disease (dACLD) groups.

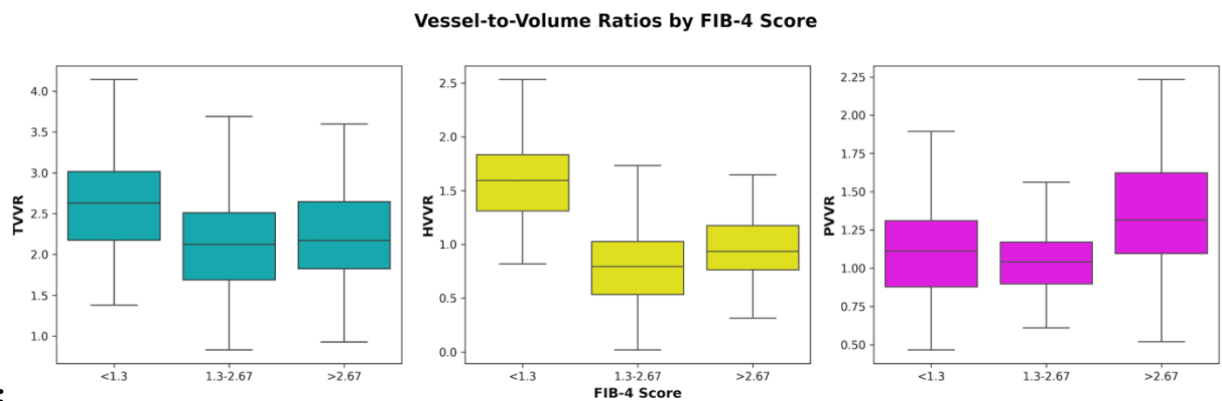

**Fig. S2:**

Fig. S2: Distribution of vessel-to-volume ratios across FIB-4 score categories. Box plots show total vessel-to-volume ratio (TVVR), hepatic vein-to-volume ratio (HVVR), and portal vein-to-volume ratio (PVVR) across three FIB-4 score categories: <1.3 (low risk), 1.3-2.67 (intermediate risk), and >2.67 (high risk) for advanced fibrosis.

## Declarations

**Competing interests**

A.B.: honoraria for consulting from Bayer, speaker fees from Bayer and Siemens. T.R. received grant support from Abbvie, Boehringer-Ingelheim, Gilead, Intercept, MSD, Myr Pharmaceuticals, Philips Healthcare, Pliant, Siemens and W. L Gore & Associates; speaking honoraria from Abbvie, Gilead, Intercept, Roche, MSD, W. L Gore & Associates; consulting/advisory board fees from Abbvie, Bayer, Boehringer-Ingelheim, Gilead, Intercept, MSD, Siemens; and travel support from Abbvie, Boehringer-Ingelheim, Gilead and Roche. M.M. received grants from Echosens, served as a speaker and/or consultant and/or advisory board member for AbbVie, Collective Acumen, Echosens, Gilead, Ipsen, Takeda, and W. L. Gore & Associates, and received travel support from AbbVie and Gilead. B.S. received travel support by AbbVie, Gilead, and Falk. G.S. received travel support from Amgen. M.T. received grant support from Albireo, Alnylam, Cymabay, Falk, Genentech, Gilead, Intercept, MSD, Takeda and UltraGenyx, honoraria for consulting from AbbVie, Albireo, Agomab, Boehringer Ingelheim, BiomX, Chemomab, Dexoligo Therapeutics, Falk, Genfit, Gilead, GSK, Hightide, Intercept, Ipsen, Jannsen, Mirum, MSD, Novartis, Phenex, Pliant, Rectify, Regulus, Siemens and Shire, speaker fees from Albireo, Boehringer Ingelheim, Bristol-Myers Squibb, Falk, Gilead, Ipsen, Intercept, MSD and Madrigal, as well as travel support from AbbVie, Falk, Gilead, Jannsen and Intercept. He is also co-inventor (service inventions as employee) of patents on the medical use of 24-norursodeoxycholic acid filed and owned by the Medical University of Graz. B.W. has given scientific presentations for Philips GmbH for which monetary compensation was received. G.L. reports a relationship with Siemens Healthineers, Boehringer Ingelheim GmbH and Roche that includes speaking and lecture fees, and a relationship with Contextflow GmbH that includes equity or stocks. G.L. and D.S. received funding by Novartis Pharmaceuticals Corporation. A.H., S.P.L., N.B., L.B., S.B., M.W.: nothing to disclose.

**Funding**

This work has been partly funded by the European Union's Horizon Europe research and innovation programme under grant agreement No. 101080302 AI-POD and No. 101136299 ARTEMIS.

**Authors contributions**

Guarantors of integrity of entire study, A.H., D.S., A.B., G.L.; study concepts/study design or data acquisition or data analysis/interpretation, all authors; manuscript drafting or manuscript revision for important intellectual content, all authors; approval of final version of submitted manuscript, all authors; agrees to ensure any questions related to the work are appropriately resolved, all authors; literature research, A.H., D.S., N.B., L.B., T.R., M.M., A.B.; clinical studies, A.H., D.S., N.B., L.B., M.M., S.P.L., G.S., B.S., T.R., A.B.; statistical analysis, A.H., D.S., L.B., M.W., first manuscript draft, A.H., and manuscript editing, all authors.

**Acknowledgements**

We would like to acknowledge Dr. Michael Weber for his outstanding support on the statistical analyses of this study.